\documentclass[reprint,superscriptaddress,amsmath,amssymb,prd,nofootinbib,longbibliography]{revtex4-2}

\usepackage{amssymb,amsmath,epsfig}
\usepackage{mathrsfs}
\usepackage{mathrsfs}
\usepackage[usenames,dvipsnames]{color}
\usepackage[bookmarks]{hyperref}
\usepackage[svgnames]{xcolor}
\usepackage{subfigure}
\usepackage{times}
\usepackage{yhmath}
\usepackage{braket}
\usepackage{soul}

\newcommand{\re}{\mbox{Re}}

\newtheorem{lemma}{Lemma}

\begin{document}

\title{Relativistic dissipative fluids in the trace-fixed particle frame: Hyperbolicity, causality, and stability}

\author{J. F\'elix Salazar}
\address{Departamento de Matem\'aticas Aplicadas y Sistemas, Universidad Aut\'onoma Metropolitana-Cuajimalpa, 05348 Cuajimalpa de Morelos, Ciudad de M\'exico M\'exico.}

\author{Ana Laura Garc\'ia-Perciante}
\address{Departamento de Matem\'aticas Aplicadas y Sistemas, Universidad Aut\'onoma Metropolitana-Cuajimalpa, 05348 Cuajimalpa de Morelos, Ciudad de M\'exico M\'exico.}

\author{Olivier Sarbach}
\address{Departamento de Matem\'aticas Aplicadas y Sistemas, Universidad Aut\'onoma Metropolitana-Cuajimalpa, 05348 Cuajimalpa de Morelos, Ciudad de M\'exico M\'exico.}
\address{Instituto de F\'isica y Matem\'aticas,
Universidad Michoacana de San Nicol\'as de Hidalgo,
Edificio C-3, Ciudad Universitaria, 58040 Morelia, Michoac\'an, M\'exico.}

\begin{abstract}
We propose a first-order theory of relativistic dissipative fluids in the trace-fixed particle frame, which is similar to Eckart's frame except that the temperature is determined by fixing the trace of the stress-energy tensor. Our theory is hyperbolic and causal provided a single inequality holds. For low wave numbers, the expected damped modes in the shear, acoustic, and heat diffusion channels are recovered. Stability of global equilibria with respect to all wave numbers is also analyzed. The conditions for hyperbolicity, causality and stability are satisfied for a simple gas of hard spheres or disks.
\end{abstract}

\date{\today}


\maketitle

\section{Introduction}
\label{Sec:Introduction}

Relativistic dissipative hydrodynamics plays a prominent role in many current problems in physics, including the description of cosmological fluids in the Universe~\cite{Maartens1996,BrO2017}, the modeling of accretion disks~\cite{Abramowicz2013,ChR2021,EHTC2022} and the study of quark-gluon plasmas encountered under extreme laboratory conditions~\cite{Shen2020}. In order to address these problems, one requires a theory which is physically sound, i.e. one that is described by hyperbolic evolution equations with causal propagation and for which equilibrium states are stable. There are many different proposals~\cite{Israel1976,Liu1986,Geroch1990,Tsumura2008,Peralta-Ramos:2009,VB2012,Kovtun2012} and a vast literature on this subject (see, for instance, \cite{Van2020,SZ2020,Gavassino2021,RDNR2024} for reviews). In particular, physically sound theories which are second-order in the off-equilibrium quantities have been developed in~\cite{Liu1986,Geroch1990,lLoSmR18}. Recently, there has been a vivid interest in theories, referred to as Bemfica-Disconzi-Noronha-Kovtun (BDNK)~\cite{BDN2018,BDN2019,pK19,Hoult2020,Bemfica2022,Kovtun2022,Rocha2022,Disconzi2024}, which are first order in the gradient expansion of the state variables. Instead of taking the entropy principle as a starting point~\cite{Liu1986,Boillat1997,CJT2014}, BDNK assume general couplings between the nonequilibrium quantities and the gradients of the state variables. Provided the coupling constants satisfy an ample set of nontrivial inequalities, a physically sound theory in the aforementioned sense is obtained.

In this article we present a formalism similar to BDNK. However, in contrast to this theory, our approach is based on the use of a specific frame, namely the trace-fixed particle frame, in which the state variables $(n,T,u^\mu)$ (particle number density, temperature parameter and mean-particle velocity) are fixed using the current density vector and the trace of the stress-energy tensor. This particular choice can be motivated by means of a kinetic (microscopic) formalism, whose details will be published elsewhere~\cite{cGaGoS25}, and it presents several advantages as we explain in the following. 

Our theory, which describes a simple non-degenerate dissipative fluid propagating on a curved spacetime and electromagnetic background, is hyperbolic and causal, provided the \emph{single} inequality~(\ref{Eq:FinalHypoCausalBound}) is satisfied. Furthermore, we establish general conditions which guarantee that, for large enough values of the only free parameter $\Lambda_0$, global equilibrium states in flat spacetime are stable with respect to modes of arbitrary wave numbers. Additionally, the known propagation of modes in the shear, acoustic and heat channels with low wave numbers is recovered independently of the value of $\Lambda_0$. We verify the fulfillment of the fundamental inequality~(\ref{Eq:FinalHypoCausalBound}) and the stability conditions for all temperatures in the case of a simple gas of hard spheres or disks in three and two dimensions, respectively.

We work on a fixed, globally hyperbolic and time-oriented $(d+1)$-dimensional spacetime with $d\geq 2$. Greek indices $\mu,\nu,\ldots$ run over $0,1,\dots,d$ and $\nabla$ denotes the Levi-Civita connection associated with the spacetime metric $g_{\mu\nu}$. $F^{\mu\nu}$ refers to the background electromagnetic field and $q$ to the charge of the fluid constituents. We use geometrized units and the signature convention $(-,+,\ldots,+)$ for the metric.

\section{Fluid equations}
\label{Sec:FluidEquations}

The equations of motions for a relativistic charged fluid are given by 
\begin{equation}
\nabla_{\mu}J^{\mu}=0,\qquad
\nabla_\mu T^{\mu\nu} + q J_\mu F^{\mu\nu} = 0,
\label{Eq:Conservations}
\end{equation}
and in this article we work with a current density and stress-energy tensor which have the form
\begin{align}
J^{\mu} &= nu^{\mu},
\label{Eq:Current}\\
T_{\mu\nu} &= \left(ne+\epsilon\right)u_{\mu}u_{\nu}
 + \left(p+\frac{\epsilon}{d} \right)\Delta_{\mu\nu}
 + 2u_{(\mu}\mathcal{Q}_{\nu)}
  + \mathcal{T}_{\mu\nu}.
\label{Eq:Stress-Energy}
\end{align}
Here, $\Delta_{\mu\nu} := g_{\mu\nu} + u_\mu u_\nu$, $u^\mu u_\mu = -1$, and $(\,)$ denotes symmetrization. The internal energy density per particle $e$ is assumed to be a function of $T$ only, and the pressure is determined through the ideal gas equation of state $p = n k_B T$, where $k_B$ is the Boltzmann constant. Further, $\epsilon$, $\mathcal{Q}^\mu$ (which is orthogonal to $u^\mu$), and $\mathcal{T}_{\mu\nu}$ (which is symmetric, trace-free and orthogonal to $u^\mu$) are off-equilibrium corrections. In particular, $\mathcal{Q}^\mu$ describes the heat flux and $\mathcal{T}_{\mu\nu}$ the trace-free part of the viscosity tensor. Meanwhile, the scalar nonequilibrium contribution $\epsilon$ accounts for the effects of the bulk viscosity, as we will see shortly.

The choice of frame (i.e. the assignment of $n$, $T$, and $u^\mu$ to a nonequilibrium state described by $J^\mu$ and $T_{\mu\nu}$) here adopted corresponds to (i) a particle frame, which fixes $n$ and $u^\mu$ such that they are related to the current density through Eq.~(\ref{Eq:Current}), (ii) the trace-fixed condition,\footnote{The unique determination of the temperature from the trace requires the condition $c_v < d k_B$ which is satisfied for a simple relativistic gas.} which determines $T$ through the trace of Eq.~(\ref{Eq:Stress-Energy}), i.e. $T^{\mu}{}_{\mu}=-ne+dp$. In contrast, Eckart's frame requires (i) and determines the temperature by fixing the internal energy through $u_{\mu}u_{\nu}T^{\mu\nu}=ne$ instead of (ii).

The nonequilibrium quantities are determined by the following first-order constitutive relations:
\begin{align}
\epsilon&=-\frac{\zeta}{\left(\frac{k_B}{c_{v}}-\frac{1}{d}\right)^{2}}\left[ \frac{\dot{T}}{T} + \frac{\theta}{d} - \Gamma_1\left( \frac{\dot{T}}{T} + \frac{k_B}{c_v}\theta \right) \right],
\label{Eq:ConstitutiveRelations_e}\\
\mathcal{Q}_{\mu} &= -\kappa\left[\frac{D_{\mu}T}{T} + a_{\mu} -
 \Gamma_{2}\left(a_{\mu}+\frac{1}{nh}D_{\mu}p-\frac{q}{h}E_{\mu}\right) \right],
\label{Eq:ConstitutiveRelations_q}\\
\mathcal{T}_{\mu\nu} &= -2\eta\sigma_{\mu\nu},
\label{Eq:ConstitutiveRelations_Pi}
\end{align}
where $\dot{T} := u^\mu\nabla_\mu T$ and $D_\mu T := \Delta_\mu{}^\nu\nabla_\nu T$. Also, $\theta := \nabla_\mu u^\mu$, $\sigma_{\mu\nu} := (\Delta_\mu{}^\alpha\Delta_\nu{}^\beta - d^{-1}\Delta_{\mu\nu}\Delta^{\alpha\beta})\nabla_{(\alpha} u_{\beta)}$ and $a^\mu = \dot{u}^\mu := u^\alpha\nabla_\alpha u^\mu$ denote the expansion, shear and acceleration, respectively. Further, $E_\mu := F_{\mu\nu} u^\nu$, $c_v:=\frac{\partial e}{\partial T}$ and $h := e + k_B T$ refer to the electric field measured by comoving observers, the heat capacity at constant volume and the enthalpy per particle. Finally, $\zeta$, $\eta$ and $\kappa$ denote the bulk and shear viscosities and the thermal conductivity which are  strictly positive.

In Eqs.~(\ref{Eq:ConstitutiveRelations_e}) and (\ref{Eq:ConstitutiveRelations_q}), $\Gamma_{1}$ and $\Gamma_{2}$ are two free functions of the temperature which are introduced through the additional freedom of adding the following combinations in the constitutive relations:
\begin{equation}
\frac{\dot{T}}{T} + \frac{k_B}{c_{v}}\theta,\qquad
\dot{u}_\mu + \frac{1}{nh}D_\mu p-\frac{q}{h}E_{\mu}.
\end{equation}
As a consequence of the balance equations~(\ref{Eq:Conservations}) these combinations are second-order in derivatives and thus Eqs.~(\ref{Eq:ConstitutiveRelations_e}) and (\ref{Eq:ConstitutiveRelations_q})
remain consistently first order for any choice of $\Gamma_1$ and $\Gamma_2$. The particular form of Eqs.~(\ref{Eq:ConstitutiveRelations_e})-(\ref{Eq:ConstitutiveRelations_Pi}) can be obtained by performing first-order transformations of frame as described in Ref.~\cite{Kovtun2019}, starting from the constitutive relations in the Eckart frame, see the companion paper~\cite{aGjSoS2024b} for details.

In the following sections the hyperbolicity, causality and stability of the system of equations given by Eqs.~(\ref{Eq:Conservations})-(\ref{Eq:ConstitutiveRelations_Pi})
is analyzed when linearized at a global equilibrium configuration, and the dependency of these properties on $\Gamma_{1}$ and $\Gamma_{2}$ is discussed. The hyperbolicity and causality of the full nonlinear system are analyzed in~\cite{aGjSoS2024b}.

\section{Hyperbolicity and causality}
\label{Sec:Hyperbolicity}

Whereas the choice $\Gamma_1=\Gamma_2=1$ eliminates the time derivatives in the constitutive relations, for the following analysis we assume $\Gamma_1,\Gamma_2\neq 1$ and use Eqs.~(\ref{Eq:ConstitutiveRelations_e}) and (\ref{Eq:ConstitutiveRelations_q}) as evolution equations for the temperature $T$ and the velocity $u^\mu$. With the help of these equations, one can eliminate $\dot{T}$ and $a_\mu$ in Eq.~(\ref{Eq:Conservations}). Together with Eq.~(\ref{Eq:ConstitutiveRelations_Pi})
one obtains the following evolution system for the fields $n$, $T$, $u^\mu$, $\epsilon$ and $\mathcal{Q}^\mu$:
\begin{eqnarray}
\dot{n} &=& -\theta n,
\label{Eq:ndot}\\
\dot{T} &=& \alpha_4\theta T
 + \alpha_5\epsilon T,
\label{Eq:Tdot}\\
\dot{u}^\mu &=& \beta_1\frac{D^\mu n}{n} + \beta_2\frac{D^\mu T}{T} + \beta_3 \mathcal{Q}^\mu -\frac{q\beta_1}{k_B T} E^\mu,
\label{Eq:udot}\\
\dot{\epsilon} &=& -D_\mu \mathcal{Q}^\mu + f_\epsilon,
\label{Eq:Edot}\\
\dot{\mathcal{Q}}^\mu &=&
2\eta D_\nu\sigma^{\mu\nu} 
 - \frac{1}{d} D^\mu\epsilon
 + f_\mathcal{Q}^\mu,
\label{Eq:qdot}
\end{eqnarray}
where
\begin{align}
& \alpha_4 := \frac{\frac{1}{d} -\frac{k_B}{c_v}\Gamma_1}{\Gamma_1-1},\qquad
\alpha_5 := \frac{1}{\Gamma_1-1}\left( \frac{k_B}{c_v}-\frac{1}{d} \right)^2\frac{1}{\zeta},
\\
& \beta_1 := -\frac{\Gamma_2\frac{k_B T}{h}}{\Gamma_2-1},\quad \beta_2 := \frac{1 - \frac{\Gamma_2 k_B T}{h}}{\Gamma_2-1},\quad
\beta_3 := \frac{1}{\Gamma_2-1}\frac{1}{\kappa},
\end{align}
and
\begin{widetext}
 \begin{eqnarray}
f_\epsilon &:=&
 -\frac{d+1}{d}\theta\epsilon + 2\eta\sigma^{\mu\nu}\sigma_{\mu\nu} - 2a_\mu\mathcal{Q}^\mu 
  + \frac{n T c_v}{\Gamma_1-1}\left( \frac{k_B}{c_v} - \frac{1}{d} \right)\left[ \theta - \left( \frac{k_B}{c_v} - \frac{1}{d} \right)\frac{\epsilon}{\zeta} \right],
\\
f_\mathcal{Q}^\mu &:=&
 2\sigma^{\mu\nu} (D_\nu + a_\nu)\eta 
 - \frac{d+1}{d} a^\mu\epsilon
 - \theta\mathcal{Q}^\mu - (\nabla^\nu u^\mu)\mathcal{Q}_\nu
 + \frac{n h}{\Gamma_2 - 1}\left[ \frac{k_B T}{h}\frac{D^\mu n}{n} - \frac{e}{h}\frac{D^\mu T}{T} - \frac{q}{h} E^\mu - \frac{\mathcal{Q}^\mu}{\kappa} \right].
\end{eqnarray}   
\end{widetext}
In the last two expressions, it is understood that $a^\mu$ is replaced with the right-hand side of Eq.~(\ref{Eq:udot}).

Equations~(\ref{Eq:ndot}--\ref{Eq:qdot}) constitute an evolution system for the fields $(n,T,u^\mu,\epsilon,\mathcal{Q}^\mu)$ which is first order in time and mixed first and second order in space derivatives (second-order spatial derivatives of the velocity field $u^\mu$ appear on the right-hand side of Eq.~(\ref{Eq:qdot})). We analyze its hyperbolic character using linearization and the principle of frozen coefficients~\cite{Kreiss89}. Accordingly, it is sufficient to linearize the equations around a homogenous background in Minkowski spacetime, such that the background solution has $n$ and $T$ constant, $(u^\mu) = (1,\vec{0})$, and $\epsilon$, $\mathcal{Q}^\mu$, $a^\mu$, $\theta$, and $\sigma_{\mu\nu}$ vanishing. Consequently, to linear order, $(u^\mu)=(1,\vec{v})$ and $(\mathcal{Q}^\mu)=(0,\vec{q})$. Furthermore, since the electric field appears as a source term, it plays no essential role for the subsequent analysis; hence from now on we set it to zero. Linearizing Eqs.~(\ref{Eq:ndot}--\ref{Eq:qdot}) and denoting by $\hat{n},\hat{T},\hat{\vec{v}},\hat{\epsilon},\hat{\vec{q}}$ the spatial Fourier transform of the perturbed fields, which are functions of the wave vector $\vec{k}$, one obtains a first-order linear system for the variables
\begin{equation}
\mathcal{N} := i k\frac{\hat{n}}{n},\quad
\mathcal{T} := i k\frac{\hat{T}}{T},
\quad
\vec{w} := i k\hat{\vec{v}},
\end{equation}
$\hat{\epsilon}$ and $\hat{\vec{q}}$, where $k := |\vec{k}|$. This system decouples into two subsystems, involving the components $(\vec{q}_\perp,\vec{w}_\perp)$ and $(q_{\|},w_{\|})$ orthogonal and parallel to $\vec{k}$.

The first system governs the propagation of the transverse modes and reads
\begin{equation}
\dot{U}_\perp = i k M_\perp U_\perp + N_\perp U_\perp,\qquad
U_\perp := \left( \begin{array}{c}
\vec{q}_\perp \\ \vec{w}_\perp
\end{array} \right),
\label{Eq:Perp}
\end{equation}
with the matrices
\begin{equation}
M_\perp := \left( \begin{array}{cc}
0 & \eta \\ \beta_3 & 0
\end{array} \right),\quad
N_\perp := \left( \begin{array}{cc}
-n h\beta_3 & 0 \\ 0 & 0
\end{array} \right).
\label{Eq:MNPerp}
\end{equation}
Hyperbolicity and causality require the matrix $M_\perp$ to be diagonalizable and to have eigenvalues which are real and smaller than one in magnitude, which yields
\begin{equation}
0 < \eta\beta_3 = \frac{1}{\Gamma_2-1}\frac{\eta}{\kappa}\leq 1.
\label{Eq:TransverseModes}
\end{equation}
In the nonrelativistic limit $T\to 0$, one can show (for instance using hard spheres or disks) that $\eta/\kappa\sim m/(k_B T)$ and thus, in order to keep $\eta\beta_{3}$ bounded, $\Gamma_2-1$ needs to increase at least as $1/T$. The fact that in this limit $h\to m$ motivates the following choice:
\begin{equation}
\Gamma_2 := \frac{h}{k_B T},
\label{Eq:Gamma2}
\end{equation}
which implies $\beta_2=0$ and $\Gamma_2 - 1 = e/(k_B T)$, such that condition~(\ref{Eq:TransverseModes}) reduces to
\begin{equation}
0 < \frac{k_B T}{e}\frac{\eta}{\kappa}\leq 1.
\label{Eq:HypoCausalBound}
\end{equation}

The second system governs the propagation of longitudinal modes $U_{\|} := \left( \mathcal{N}, \mathcal{T}, q_{\|}, w_{\|}, \hat{\epsilon} \right)^T$ and has the form
\begin{equation}
\dot{U}_{\|} = i k M_{\|} U_{\|} 
 + N_{\|} U_{\|}.
\label{Eq:LongitudinalLin}
\end{equation}
Here, $M_{\|}$ is a $5\times 5$ matrix which has the block form 
$M_{\|} = \left( \begin{array}{cc}
0 & Q \\ R & 0
\end{array} \right)$,
with the matrices $Q$ and $R$ given by
\begin{equation}
Q := \left( \begin{array}{cc}
-1 & 0 \\
\alpha_4 & \alpha_5 \\
2\frac{d-1}{d}\eta & -\frac{1}{d}
\end{array} \right),\quad
R := \left( \begin{array}{ccc}
\beta_1 & \beta_2 & \beta_3 \\
0 & 0 & -1
\end{array} \right),
\end{equation}
and the exact form of $N_{\|}$ will not be needed for the moment. The block structure of $M_{\|}$ allows one to analyze its eigenvalues using the following lemma whose proof can be found in Lemma~1 of Ref.~\cite{oSeBjP19}:

\begin{lemma}
Suppose the $2\times 2$ matrix $RQ$ is diagonalizable and has nonzero eigenvalues $\mu_1$ and $\mu_2$. Then, the matrix $M_{\|}$ is diagonalizable and its eigenvalues are
\begin{equation}
0,\pm\sqrt{\mu_1},\pm\sqrt{\mu_2}.
\end{equation}
\end{lemma}

Therefore, the system~(\ref{Eq:LongitudinalLin}) is hyperbolic and causal if the eigenvalues $\mu_1$ and $\mu_2$ of $RQ$ satisfy $0 < \mu_1 < \mu_2\leq 1$. In terms of the trace $Tr$ and determinant $D$ of $RQ$, it is simple to check that these conditions are equivalent to
\begin{equation}
0 < D < 1,\quad
2\sqrt{D} < Tr \leq 1 + D.
\label{Eq:HypoCausalCond}
\end{equation}
For the choice~(\ref{Eq:Gamma2}) one finds
\begin{equation}
D = \frac{1}{d}\frac{k_B T}{e},\qquad
Tr = \frac{1}{d} + \frac{k_B T}{e}\left[ 1 + 2\frac{d-1}{d}\frac{\eta}{\kappa} \right].
\label{Eq:DT}
\end{equation}
For the second condition in Eq.~(\ref{Eq:HypoCausalCond}) one first notes that
\begin{equation}
Tr > \frac{1}{d} + \frac{k_B T}{e} \geq 2\sqrt{\frac{1}{d}\frac{k_B T}{e}} = 2\sqrt{D},
\end{equation}
and next that
\begin{equation}
1 + D - Tr = \left( 1 - \frac{1}{d} \right)\left[ 1 - \frac{k_B T}{e}\left( 1 + 2\frac{\eta}{\kappa} \right) \right],
\end{equation}
which implies that it is equivalent to
\begin{equation}
\frac{k_B T}{e}\left(1 + 2\frac{\eta}{\kappa}\right) \leq 1.
\label{Eq:FinalHypoCausalBound}
\end{equation}
In particular, this inequality entails the validity of $D < 1$ and of the upper bound in Eq.~(\ref{Eq:HypoCausalBound}) for the transverse modes. 

Summarizing, the system~(\ref{Eq:ndot}--\ref{Eq:qdot}) is hyperbolic and causal as long as $T$, $e$, $\eta$, and $\kappa$ are strictly positive and satisfy the fundamental inequality~(\ref{Eq:FinalHypoCausalBound}). In the companion paper~\cite{aGjSoS2024b} we prove that these same conditions lead to the strong hyperbolicity of the full nonlinear system and to a well-posed Cauchy problem provided $\eta\beta_3$, $\mu_1$ and $\mu_2$ are distinct.

\section{Fundamental inequality and characteristic speeds for hard disks and spheres}
\label{Sec:Hardspheres}

\begin{figure}[htb]
\includegraphics[scale=0.9]{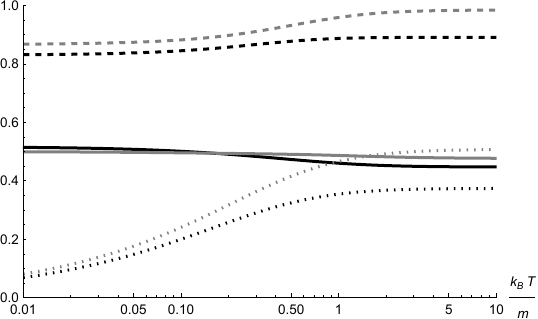}
\caption{Characteristic speeds for hard spheres (black) and hard disks (gray). The solid, dashed and dotted lines refer to the speeds $\sqrt{\beta_3\eta}$, $\sqrt{\mu_1}$ and $\sqrt{\mu_2}$, respectively. The limits for $T\to 0$ and $T\to \infty$ can again be understood from the expressions provided in Appendix~\ref{App:A}.}
\label{Fig:beta3eta}
\end{figure}

Next, we verify the validity of the fundamental inequality~(\ref{Eq:FinalHypoCausalBound}) and compute the characteristic speeds for a simple gas of hard disks or spheres \footnote{It is also interesting to notice that for the Marle model in the relaxation time approximation for a simple gas in $d$ space dimensions, one finds~\cite{CercignaniKremer-Book,cGaGoS25}
$$
\frac{\eta}{\kappa} = \frac{h}{(c_v + k_B)T}.
$$
The left-hand side of Eq.~(\ref{Eq:FinalHypoCausalBound}) converges to $4/(d+2)$ and $3/d$ in the limits $T\to 0$ and $T\to \infty$, respectively. Therefore, within this approximation, the fundamental inequality~(\ref{Eq:FinalHypoCausalBound}) is satisfied for all temperatures in $d\geq 3$ dimensions only.
}.
Known explicit expressions for the transport coefficients for these models can be found in~\cite{almaana2D2019,CercignaniKremer-Book} and are summarized in Appendix~\ref{App:A} for convenience. Figure~\ref{Fig:Fundamental} shows the left-hand side of the fundamental inequality~(\ref{Eq:FinalHypoCausalBound}), from which it is clear that the inequality is satisfied for all temperatures. Figure~\ref{Fig:beta3eta} shows the behavior of the nonzero characteristic speeds $\sqrt{\beta_3\eta}$, $\sqrt{\mu_1}$, and $\sqrt{\mu_2}$ as a function of the temperature, where $\beta_3\eta = \frac{k_B T}{e}\frac{\eta}{\kappa}$ and $\mu_{1,2} = (Tr \pm \sqrt{Tr^2 - 4D})/2$ can be computed from Eq.~(\ref{Eq:DT}). Note that for spheres these speeds are always different from each other whereas for disks there is a crossing of the speeds $\sqrt{\beta_3\eta}$ and $\mu_2$ at a temperature $T\sim m/k_B$.

\begin{figure}[htb]
\includegraphics[scale=0.9]{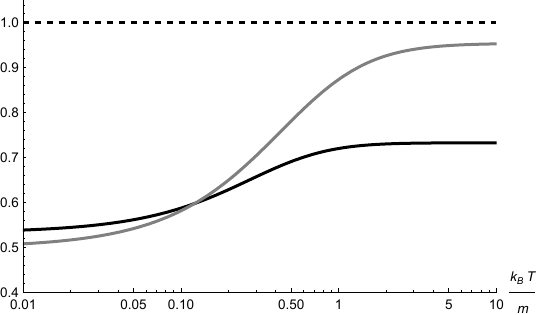}
\caption{Left-hand side of Eq.~(\ref{Eq:FinalHypoCausalBound}) for hard spheres (black) and hard disks (gray). The horizontal dashed line, which is the upper bound of the inequality, is shown for reference. For $T\to 0$ one finds that this expression converges to $8/15 = 0.533\ldots$ for $d=3$ and to $1/2$ for $d=2$, whereas for $T\to \infty$ it converges to $11/15 = 0.733\ldots$ for $d=3$ and to $21/22=0.9545\ldots$ for $d=2$ (see Appendix~\ref{App:A}).}
\label{Fig:Fundamental}
\end{figure}

\section{Stability and sound propagation}
\label{Sec:Stability}

Next, we analyze the behavior of the solutions for wavelengths $\lambda$ satisfying $\ell_{mfp}\ll \lambda\ll \ell_{ms}$, with $\ell_{mfp}$ and $\ell_{ms}$ denoting the particles' mean free path and the macroscopic scale, respectively. In this regime, it is possible to use the first-order theory and, at the same time, to freeze the coefficients as we have done in the previous section. It is illustrative to keep $\Gamma_1$ and $\Gamma_2$ arbitrary for the following analysis.

For transverse modes, one obtains from Eqs.~(\ref{Eq:Perp}) and (\ref{Eq:MNPerp}) solutions of the form $U_\perp(t) = e^{st} U_\perp(0)$ with $s = s_\pm$ and
\begin{equation}
s_\pm = \frac{n h\beta_3}{2}\left[ -1 \pm \sqrt{ 1 - \frac{4\eta k^2}{n^2 h^2\beta_3}} \right].
\end{equation}
Since $\eta/(nh),\,\kappa/(nh)\sim \ell_{mfp}$ one can expand the square root and obtain $s_+ = -\frac{\eta k^2}{n h} + {\cal O}(\ell_{mfp}^3 k^4)$, whereas $s_- = -\frac{1}{\Gamma_2-1}\frac{nh}{\kappa} + {\cal O}(\ell_{mfp} k^2)$. Provided $\Gamma_2 > 1$ (as required by hyperbolicity) the mode with $s_-$ is rapidly damped (on a timescale of $\ell_{mfp}$). The mode with $s_+$ has a slower damping time of the order of $1/(\ell_{mfp} k^2)$ which is independent of $\Gamma_1$ and $\Gamma_2$ and corresponds to the shear channel~\cite{Weinberg71,Groot-Book,Hoult2020}. Since $\re(s_\pm) < 0$ for all $k\neq 0$ the transverse system is stable.

Similarly, longitudinal modes are of the form $U_{\|}(t) = e^{st} U_{\|}(0)$ where $s$ is an eigenvalue of the $5\times 5$ matrix $P_{\|} := ik M_{\|} + N_{\|}$. Explicitly, one finds
\begin{equation}
    P_\parallel = 
    \begin{pmatrix}
     0 & 0 & 0 & -ik & 0 \\
     0 & 0 & 0 & ik \alpha_4 & ik \alpha_5 \\
     H_0\frac{k_B T\kappa}{h} & -H_0 \frac{e\kappa}{h} & -H_0 & 2ik\eta\frac{d-1}{d} & -\frac{ik}{d} \\
     ik \beta_1 & ik\beta_2 & ik \beta_3 & 0 & 0 \\
     0 & 0 & -ik & H_1\frac{\zeta}{\frac{k_B}{c_v} - \frac{1}{d}} & -H_1
\end{pmatrix},
\end{equation}
where we have abbreviated $H_0 := nh\beta_3$ and $H_1 := n T c_v\alpha_5$.
After some calculations one finds that the characteristic polynomial of the matrix  $P_{\|}$ has the following structure:
\begin{equation}
\begin{split}
p(s) & := \det\left( s I - P_{\|} \right) 
 = s^3\left(s + H_0\right)\left(s + H_1 \right) \\
 & + k^2\left( c_3 s^3 + c_2 s^2 + c_1 s \right) 
 + k^4\left( d_1 s + d_0 \right),
 \end{split}
\label{Eq:CharPoly}
\end{equation}
where the coefficients $c_i$ and $d_i$ are independent of $k$ and $s$ and can be found in Appendix~\ref{App:B}.

When $k = 0$ one obtains the eigenvalues $0$ (three times degenerated), $-H_0$ and $-H_1$. If $\Gamma_1,\Gamma_2 > 1$, the last two yield modes that are rapidly damped, on timescales $\ell_{mfp}$. In order to investigate the behavior of the remaining modes for small values of $k$ we use perturbation theory (see Appendix~\ref{App:B} for details) and obtain eigenvalues of the form $s = -\frac{1}{\tau} \pm i v k + {\cal O}(\ell_{mfp}^2 k^3)$. There is a purely damped mode~\cite{Groot-Book,Hoult2020}, describing heat diffusion, for which 
\begin{equation}
\frac{1}{\tau} = \frac{\kappa k^2}{n c_p T},\qquad v=0,
\label{Eq:DampedHeatDiffusion}
\end{equation}
and two oscillating damped acoustic modes~\cite{Weinberg71,Groot-Book,Hoult2020} for which
\begin{equation}
\frac{1}{\tau} =
\frac{k^2}{2nh}\left[ A + v^{2}\kappa\left(\frac{h}{c_p T}-1\right)^2\right],\quad
v = \sqrt{\frac{k_B T}{h}\frac{c_p}{c_v}},
\label{Eq:AcousticDampingITime}
\end{equation}
where $A := 2\left(1-\frac{1}{d}\right)\eta + \zeta$ is the Stokes attenuation coefficient and $c_p := c_v + k_B$ the specific heat at constant pressure per particle. Therefore, the expected dynamics for low wave numbers is recovered.

Stability of the equilibrium configuration requires that $\re(s) < 0$ for all $k\neq 0$. Based on the Routh-Hurwitz criterion~\cite{RouthHurwitzTest1,RouthHurwitzTest2} one can prove that this is achieved, under the assumptions (i)-(iv) detailed in Appendix~\ref{App:B}, by setting
\begin{equation}
\Gamma_1 = 1 + \frac{c_v}{k_B}\frac{e^2}{k_B T h}\left(\frac{k_B}{c_{v}}-\frac{1}{d}\right)^2\frac{\kappa}{\zeta}\Lambda_0, \label{Eq:Gamma1}
\end{equation}
with large enough values of the constant $\Lambda_0$ and the choice~(\ref{Eq:Gamma2}) for $\Gamma_2$. In particular, the assumptions (i)-(iv) are satisfied for a simple gas of hard spheres or disks.

\section{Conclusions}
\label{Sec:Conclusions}

In this work we proposed a new first-order theory of relativistic dissipative fluids based on the trace-fixed particle frame. We showed that this theory is hyperbolic and causal provided the single inequality~(\ref{Eq:FinalHypoCausalBound}) is satisfied. Moreover, we have specified sufficient conditions for global equilibria to be stable. Although similar in spirit to BDNK, our constitutive relations~(\ref{Eq:ConstitutiveRelations_e}--\ref{Eq:ConstitutiveRelations_Pi}) are \emph{different} from the ones in~\cite{Bemfica2022}, as explained in the companion article~\cite{aGjSoS2024b}. In particular, the use of the trace-fixed particle frame, together with the freedom of choosing the temperature-dependent functions $\Gamma_1$ and $\Gamma_2$ according to Eqs.~(\ref{Eq:Gamma2}) and (\ref{Eq:Gamma1}), naturally leads to a physically sound theory. The implications of the inequality~(\ref{Eq:FinalHypoCausalBound}) and the stability conditions need to be further explored. However, we have shown that there is at least one relevant case in which they are satisfied for all temperatures, namely for a simple gas consisting of hard disks or spheres. Furthermore, our theory correctly reproduces the known propagation of modes in the shear, acoustic and heat channels with low wave numbers. Finally, as shown in~\cite{aGjSoS2024b}, our theory also implies positive entropy production within its limit of validity. For simplicity we have restricted this work to the case of a fixed background spacetime, and thus it only applies to physical situations in which the self-gravity of the fluid can be neglected. Future work will analyze the Cauchy problem for the full coupled Einstein-fluid equations. Therefore, our formalism should serve as a starting point for a systematic study of transport phenomena in relativistic astrophysics, cosmology and high-energy physics.

\acknowledgments

It is a pleasure to thank Luis Lehner, Oscar Reula, and Thomas Zannias for fruitful discussions. We also thank Luis Lehner for comments on a previous version of this manuscript. O.S. was partially supported by CIC Grant No.~18315 to Universidad Michoacana and by CONAHCyT Network Project No.~376127 ``Sombras, lentes y ondas gravitatorias generadas por objetos compactos astrof\'isicos". FS was supported by a CONAHCyT postdoctoral fellowship.

\appendix
\section{Transport coefficients for hard disks and spheres}
\label{App:A}

Here we summarize relevant formulas for a simple relativistic gas in $d=2,3$ space dimensions with a hard disk/sphere interaction model. For details and derivations see~\cite{CercignaniKremer-Book} for $d=3$ and \cite{almaana2D2019} for $d=2$.

First, recall that the enthalpy per particle is given by~\cite{rAcGoS2022}
\begin{equation}
h = m\mathcal{G}(z),\qquad
\mathcal{G}(z) := \frac{K_{\frac{d+3}{2}}(z)}{K_{\frac{d+1}{2}}(z)},
\end{equation}
where here and in the following, $m$ denotes the mass of the particles, $z := m/(k_B T)$ and $K_\nu$ refer to the modified Bessel functions of the second type. Accordingly, the heat capacities at constant pressure and volume per particle are given by
\begin{equation}
c_p = - k_B z^2\mathcal{G}'(z),\qquad
c_v = c_p - k_B,
\end{equation}
where $\mathcal{G}'$ denotes the derivative of $\mathcal{G}$. One can show that as the temperature $T$ increases from $0$ to $\infty$, $c_v$ increases monotonically from $d/2$ to $d$ and the quantity $\nu := k_B T/e$ from $0$ to $1/d$.

From Eqs.~(5.89-5.91) in~\cite{CercignaniKremer-Book} the transport coefficients for hard spheres in $d=3$ are, after translating to our notation,
\begin{align}
\zeta &= \frac{1}{64\pi}\frac{m}{z\sigma}
\left[ \frac{k_B}{c_v} \frac{d}{dz}\left( \frac{c_v}{k_B} \right) \right]^2
\frac{z^4 K_2(z)^2}{2K_2(2z) + z K_3(2z)},
\\
\eta &= \frac{15}{64\pi}\frac{m}{z\sigma}
\frac{z^4 K_3(z)^2}{(2 + 15z^2)K_2(2z) + z(49 + 3z^2)K_3(2z)},
\\
\kappa &= \frac{3}{64\pi}\frac{m}{z\sigma}
\left( \frac{c_p}{k_B} \right)^2
\frac{z^2 K_2(z)^2}{(2 + z^2)K_2(2z) + 5z K_3(2z)},
\end{align}
where $\sigma$ is the (constant) cross section. Consequently, the ratios
\begin{align}
    \frac{\zeta}{\kappa} & = \frac{1}{3}\left[ \frac{\frac{d}{dz}\left( z^2\mathcal{G}'(z) \right)}{z\mathcal{G}'(z)(1 + z^2\mathcal{G}'(z))} \right]^2\frac{2+z^2 + 5z\mathcal{G}(2z)}{2 + z\mathcal{G}(2z)}, \\
\frac{\eta}{\kappa} & = 5\left[ \frac{\mathcal{G}(z)}{z\mathcal{G}'(z)} \right]^2\frac{2 + z^2 + 5z\mathcal{G}(2z)}{2 + 15z^2 + z(49 + 3z^2)\mathcal{G}(2z)},
\end{align}
are functions of the temperature only. In the nonrelativistic limit $z\to \infty$ one finds $\mathcal{G}(z) = 1 + 5/(2z) + \mathcal{O}(1/z^2)$ and it follows that $\nu\eta/\kappa\to 4/15$ and $\nu\zeta/\kappa\to 0$, whereas in the ultrarelativistic limit, $\mathcal{G}(z) \sim 4/z$, which implies $\nu\eta/\kappa\to 1/5$ and $\nu\zeta/\kappa\to 0$. Using this information, one finds the following values for the characteristic speeds in the limit $T\to 0$:
\begin{align}
    \sqrt{\beta_3\eta} & =\frac{2}{\sqrt{15}} \simeq 0.516,\quad
\sqrt{\mu_1}=\sqrt{\frac{31}{45}}\simeq 0.830, \nonumber \\
\sqrt{\mu_2} & = 0,
\end{align}
while for $T\to\infty$:
\begin{align}
    \sqrt{\beta_3\eta} & = \frac{1}{\sqrt{5}} \simeq 0.447,\quad
\sqrt{\mu_1}=\sqrt{\frac{7+ 2\sqrt{6}}{15}}\simeq 0.891, \nonumber \\
\sqrt{\mu_2} & = \sqrt{\frac{7- 2\sqrt{6}}{15}}\simeq0.374.
\end{align}

For $d=2$ one has $\mathcal{G}(z) = (z^2 + 3z + 3)/(z(z+1))$ and for hard disks the transport coefficients are~\cite{almaana2D2019}
\begin{align}\label{kappa2D}
    \zeta & = \frac{15m}{R\mathcal{I}_{1}}\frac{1}{z^2\left(z^2+4z +2\right)^{2}},
\quad
\eta = \frac{30m}{R\mathcal{I}_{2}}\frac{\left(z^2 + 3z + 3\right)^{2}}{z^6}, \nonumber \\
\kappa & = \frac{15m}{R\mathcal{I}_{3}}\frac{(2z^2 + 6z + 3)^2}{z^6\left(1+z\right)^{2}},
\end{align}
where $R$ is the radius of the disks and
\begin{align}
\mathcal{I}_{1} &=\int_{2z}^{\infty}\left(\frac{1}{x}+\frac{3}{x^{2}}+\frac{3}{x^{3}}\right) w(x,z) dx, 
\\
\mathcal{I}_{2} &= \int_{2z}^{\infty}\left(x+2+\frac{3}{x}+\frac{3}{x^{2}}+\frac{3}{x^{3}}\right) w(x,z)dx,
\\
\mathcal{I}_{3} &= \int_{2z}^{\infty}\left(1+\frac{3}{x}+\frac{6}{x^{2}}+\frac{6}{x^{3}}\right) w(x,z) dx,
\end{align}
with $w(x,z) := \left(\frac{x^2}{z^2}-4\right)^{5/2} e^{-\left(x-2z\right)}$. Notice that, in order to obtain the thermal conductivity for the hard disks model, one needs to resort to the transformations established in Ref.~\cite{Kovtun2019} (see also Sec.~IV in Ref.~\cite{JNET24}). Indeed, since in  Ref.~\cite{almaana2D2019} the heat flux is written as (translating to the notation and signature of the present work)
\begin{equation}
\mathcal{Q}^{\mu}=-\left(L_{T}\frac{D^\mu T}{T}+L_{n}\frac{D^\mu n}{n}\right),
\end{equation}
one obtains $\kappa=\frac{h}{e}L_T$ which leads to the expression for $\kappa$ given by Eq.~(\ref{kappa2D}).

As in the three-dimensional case, the ratios $\zeta/\kappa$ and $\eta/\kappa$ are functions of the temperature only. For $z\to \infty$ one finds by means of the variable substitution $x = 2z + \xi$ that 
\begin{equation}
\mathcal{I}_{1}\sim \frac{30\sqrt{\pi}}{z^{7/2}},\qquad
\mathcal{I}_{2}\sim \frac{120\sqrt{\pi}}{z^{3/2}},\qquad
\mathcal{I}_{3}\sim \frac{60\sqrt{\pi}}{z^{5/2}},
\end{equation}
which leads to $\nu\eta/\kappa\to 1/4$  and the following values for the characteristic speeds for $T\to 0$:
\begin{equation}
\sqrt{\beta_3\eta}=\frac{1}{2},\quad\sqrt{\mu_1}=
\frac{\sqrt{3}}{2} \simeq 0.866,\quad
\sqrt{\mu_2}=0.
\end{equation}
For $z\to 0$ one finds 
\begin{equation}
\mathcal{I}_{1}\sim \frac{48}{z^5},\qquad
\mathcal{I}_{2}\sim \frac{1056}{z^5},\qquad
\mathcal{I}_{3}\sim \frac{240}{z^5},
\end{equation}
which yields $\nu\eta/\kappa\to 5/22$ and the following  characteristic speeds for $T\to \infty$:
\begin{align}
    \sqrt{\beta_3\eta} & = \sqrt{\frac{5}{22}}\simeq 0.477,\quad
\sqrt{\mu_1}=\frac{1}{2}\sqrt{\frac{27+ 7\sqrt{5}}{11}} \simeq 0.985, \nonumber \\
\sqrt{\mu_2} & = \frac{1}{2}\sqrt{\frac{27- 7\sqrt{5}}{11}}\simeq0.508.
\end{align}

\section{Characteristic polynomial and stability}
\label{App:B}

Here we provide details regarding the roots of the characteristic polynomial $p(s)$ defined in Eq.~(33), which determines the complex frequencies of the longitudinal modes:
\begin{align}\label{S:SMChaPol1}
    p(s) & = s^3(s + H_0)(s + H_1) \nonumber \\
   &+ k^2\left( c_3 s^3 + c_2 s^2 + c_1 s \right) + k^4\left( d_1 s + d_0 \right),
\end{align}
where $H_0 = nh\beta_3$ and $H_1 = n T c_v\alpha_5$. Explicitly, the coefficients $c_i$ and $d_i$ are given by
\begin{eqnarray}
d_0 &=& H_0 H_1 \frac{\kappa v_s^2}{n T c_p},
\label{S:SMCd0}\\
d_1 &=& \frac{1}{d}\left[ \kappa\beta_3 - (1-\alpha_4)\beta_2 \right] + \alpha_5\beta_2 A_0,
\label{S:SMCd1}\\
c_1 &=& H_0 H_1 v_s^2,
\label{S:SMCc1}\\
 c_2 &=& \frac{H_0 H_1}{n h}\left[ A + \kappa\left( \frac{k_B T}{h} + \frac{e^2}{h T c_v} \right)
\right] \nonumber \\
 & + & (H_0 + H_1) v_s^2,
\label{S:SMCc2}\\
c_3 &=& \frac{1}{d} + \beta_3(\kappa + A_0) - (1-\alpha_4)\beta_2,
\label{S:SMCc3}
\end{eqnarray}
where
\begin{equation}
v_s := \sqrt{ \frac{k_B T}{h}\frac{c_p}{c_v}},
\end{equation}
is the speed of sound, $A_0:=2\left(1-\frac{1}{d}\right)\eta$, and $A := A_0 + \zeta$. Note that $d_1$ and $c_3$ are equal to the determinant $D$ and the trace $Tr$ of the matrix $RQ$, respectively. This can be understood by analyzing the behavior of the roots of $p(s)$ in the high-frequency limit. Indeed, considering $s = i\lambda k$ with $k\to \infty$, leads to
\begin{equation}
\lim\limits_{k\to\infty} \frac{p(s)}{k^5} = i\lambda(\lambda^4 - c_3\lambda^2 + d_1),
\end{equation}
such that $\lambda^2$ are the eigenvalues of the matrix $RQ$. Before we proceed, we note that the coefficients $d_0$, $c_1$ and $c_2$ depend on the two paramters $\Gamma_1$ and $\Gamma_2$ of the theory only through the combinations $H_0 H_1$ and $H_0 + H_1$. This will play an important role in the following.

\subsection{Behavior of the roots for small wave numbers}

For small $|k|$ there are two roots of $p(s)$ which are of the form
\begin{equation}
s = -H_0 + {\cal O}(\ell_{mfp} k^2),\quad
s = -H_1 + {\cal O}(\ell_{mdf} k^2),
\end{equation}
and from now on we assume $\Gamma_1,\Gamma_2 > 1$ such that $H_0$ and $H_1$ are positive which implies that these roots describe rapidly damped modes, as explained in the main text. The other three roots are obtained from the ansatz $s = k^r y(k)$ with $y$ a smooth function of $k$ which converges to $y_0\neq 0$ as $k\to 0$. Substituting this ansatz into the characteristic polynomial gives
\begin{align}\label{Eq:pskr}
     p(s) & = k^{3r} y^3\left[ H_0 H_1 + (H_0 + H_1)k^r y + k^{2r} y^2 \right] \nonumber \\
 & + k^{2+r} y\left( c_1 + c_2 k^r y + c_3 k^{2r} y^2 \right)
 + k^4\left( d_0 + d_1 k^r y \right).
\end{align}
We see from this that there are two possible values for $r$, namely $r=1$ and $r=2$. In the first case one obtains
\begin{align}\label{Eq:Fky1}
   0 = \mathcal{F}(k,y) & := \frac{p(k y)}{k^3}
   = c_1 y + H_0 H_1 y^3 \nonumber \\
 & + k\left[ d_0 + c_2 y^2 +(H_0 + H_1) y^4 \right] 
+ {\mathcal O}(k^2), 
\end{align}
which in the limit $k\to 0$ yields
\begin{equation}
0 = \mathcal{F}(0,y_0) = y_0(c_1 + H_0 H_1 y_0^2),
\end{equation}
and, since $y_0\neq 0$ and $H_0 H_1 > 0$, leads to
\begin{equation}
y_0 = \pm i\sqrt{\frac{c_1}{H_0 H_1}} = \pm i v_s.
\label{Eq:y0}
\end{equation}
Since
\begin{equation}
\frac{\partial\mathcal{F}}{\partial y}(0,y_0) = c_1 + 3H_0 H_1 y_0^2 = -2c_1 < 0,
\end{equation}
the implicit function theorem guarantees, for small enough values of $|k|$, the existence of a smooth function $y(k)$ satisfying $\mathcal{F}(k,y(k)) = 0$ and $y(0) = y_0$. Differentiating both sides of Eq.~(\ref{Eq:Fky1}) with respect to $k$ and evaluating at $k=0$ yields
\begin{align}\label{Eq:yprime0}
    y'(0) & = \frac{d_0 + c_2 y_0^2 + (H_0 + H_1)y_0^4}{2c_1} \nonumber \\
 & = -\frac{1}{2nh}\left[ A + \kappa v_s^2\left( 1 - \frac{h}{T c_p} \right)^2 \right].
\end{align}
Hence, we obtain solutions for small $|k|$ of the form $s = \pm i v_s k + y'(0)k^2 + \mathcal{O}(k^3)$ which lead to the damped acoustic modes described in Eq.~(35) of the main text.

For $r=2$ one obtains from Eq.~(\ref{Eq:pskr})
\begin{align}
  0 = \mathcal{G}(k^2,y) &:= \frac{p(k^2 y)}{k^4}
  = d_0 + c_1 y \nonumber \\
 & + k^2\left[ d_1 y + c_2 y^2 + H_0 H_1 y^3 \right]
+ {\mathcal O}(k^4).  
\end{align}
For $k=0$ this yields $y_0 = -\frac{d_0}{c_1} = -\frac{\kappa}{n c_p T}$, and since
\begin{equation}
\frac{\partial\mathcal{G}}{\partial y}(0,y_0) = c_1 > 0,
\end{equation}
the implicit function theorem guarantees again the existence of a smooth local solution $y(k^2)$ of $\mathcal{G}(k^2, y(k^2)) = 0$ such that $y(0) = y_0$. This gives rise to the damped heat diffusion modes described in Eq.~(34) of the main text.

\subsection{Mode stability for all wave numbers}

To analyze the stability with respect to all wave numbers $k$, the Routh-Hurwitz criterion (see, for instance, \cite{RouthHurwitzTest1,RouthHurwitzTest2} for elementary proofs) is applied to the fifth-order polynomial Eq.~(\ref{S:SMChaPol1}) written as:
\begin{equation}
p(s) = s^5 + a_1s^4 + a_2 s^3 + a_3 s^2 + a_4 s + a_5,
\end{equation}
with $a_1 = H_0 + H_1$, $a_2 = H_0 H_1 + c_3 k^2$, $a_3 = c_2 k^2$, $a_4 = c_1 k^2 + d_1 k^4$, $a_5 = d_0 k^4$. To ensure $\re(s) < 0$ for all roots, it is essential that (i) $a_i > 0$ for all $i = 1,2,3,4,5$ and that (ii) the following inequalities are satisfied:
\begin{align}
\det(H_2) & := a_1 a_2 - a_3  > 0, \\
\det(H_3) & := a_3\det(H_2) - a_1(a_1 a_4 - a_5) > 0, \\
\det(H_4) & := a_4 \text{det}(H_3) - a_5(a_1 a_2^2 + a_5 - a_2 a_3 -a_1 a_4) > 0.
\end{align}
Explicitly, one finds
\begin{align}
\det(H_2) & = H_0 H_1(H_0 + H_1) + B_1 k^2,
\\
\det(H_3) & = (H_0 + H_1) B_2 k^2 
 + \left[ c_2 B_1 - (H_0 + H_1) B_3\right] k^4,
\\
\det(H_4) & = (H_0 + H_1)\left[c_1 B_2 - d_0(H_0 H_1)^2\right]k^4 \nonumber \\
& + \left[c_1 c_2 B_1 - 2c_1(H_0 + H_1)B_3 + c_2 H_0 H_1 B_3 \right. \nonumber \\
&\left.- 2d_0 H_0 H_1 B_1\right]k^6
    + \left(c_3 B_1 B_3 - B_3^2 - d_1 B_1^2\right)k^8,
\end{align}
with
\begin{align}
    B_1 & = (H_0 + H_1) c_3 - c_2, \quad
B_2 = H_0 H_1 c_2 - (H_0 + H_1) c_1,
\nonumber \\
B_3 & = (H_0 + H_1) d_1 - d_0.
\end{align}
In order to have $a_i > 0$ and $\det(H_i) > 0$ for all $i=2,3,4$ and $k\neq 0$ (and keeping in mind that $H_0$ and $H_1$ are strictly positive) one is led to the following set of inequalities that must be fulfilled: $d_0,d_1,c_1,c_2,c_3 > 0$,
\begin{align}
& B_1,B_2 > 0,
\label{Eq:Ineq0}\\
& D_1 := c_2 B_1 - (H_0 + H_1)B_3 > 0, 
\label{Eq:Ineq1}\\
& D_2 := c_1 B_2 - d_0(H_0 H_1)^2 > 0,
\label{Eq:Ineq2}\\
& D_3 := c_3 B_1 B_3 - B_3^2 - d_1 B_1^2 > 0,
\label{Eq:Ineq3}
\end{align}
and
\begin{align}
D_4 & := (c_1 c_2 - 2d_0 H_0 H_1)B_1 + \left[ c_2 H_0 H_1 \right. \nonumber \\
&\left. - 2c_1(H_0 + H_1)\right]B_3 > 0.
\label{Eq:Ineq4}
\end{align}
(Alternatively, the last expression could be negative and suitably bounded from below to make sure that $\det(H_4) > 0$ for all $k\neq 0$; however we will not consider this option here.)

For the following, we analyze the validity of these conditions for the particular choice $\Gamma_2 = h/(k_B T)$ which has $\beta_1 = -\nu$, $\beta_2 = 0$ and $\beta_3 = \nu/\kappa$, where we recall the definition $\nu := \frac{k_B T}{e}$. In this case one finds $H_0 = nh\nu/\kappa$ and
\begin{align}
    d_0 & = \nu\frac{H_0}{\Lambda} \frac{k_B}{c_v},
\qquad
d_1 = \frac{\nu}{d},\qquad
c_1 = \frac{H_0^2}{\Lambda}v_s^2, \nonumber \\
c_2 & = H_0 v_s^2 + \frac{H_0}{\Lambda}\left(\nu + \nu\frac{A}{\kappa}+ \frac{k_B}{c_v}\right),\quad
c_3 = \frac{1}{d} + \nu\left( 1 + \frac{A_0}{\kappa} \right),
\end{align}
where for simplicity we have set $\Lambda:=H_0/H_1$ which is positive. Explicitly, one finds
\begin{equation}
\Lambda = \frac{h}{e}\frac{k_B}{c_v}\left( \frac{k_B}{c_v} - \frac{1}{d} \right)^{-2}\frac{\zeta}{\kappa}(\Gamma_1 - 1),
\label{Eq:LambdaGamma1}
\end{equation}
such that choosing $\Gamma_1$ is equivalent to fixing $\Lambda$. In turn, the coefficients $B_i$ are given by
\begin{align}
   B_1 & = H_0\left[c_3 - v_s^2 - \frac{1}{\Lambda}\left(\frac{k_B}{c_v} - \frac{1}{d} + \nu\frac{\zeta}{\kappa}\right)\right],
\nonumber \\
B_2 & = \frac{H_0^3}{\Lambda^2}\left[\nu - v_s^2 + \frac{k_B}{c_v} + \nu\frac{A}{\kappa} \right],
\nonumber \\
B_3 & = \nu H_0 \left[\frac{1}{d} - \frac{1}{\Lambda}\left(\frac{k_B}{c_v} - \frac{1}{d}\right)\right]. 
\end{align}

For the following, we claim that when $\Lambda = \Lambda_0/\nu$ with a large enough constant $\Lambda_0$, the inequalities~(\ref{Eq:Ineq0}--\ref{Eq:Ineq4}) are satisfied provided $e(T)$ and the the transport coefficients $\zeta$, $\eta$, and $\kappa$ satisfy the following conditions:
\begin{enumerate}
\item[(i)] $e: (0,\infty)\to (m,\infty)$ is a smooth increasing function satisfying $e(T)\to m > 0$ for $T\to 0$,
\item[(ii)] The heat capacity per particle $c_v = \frac{\partial e}{\partial T}$ is positive and converges to positive values for $T\to 0$ and $T\to \infty$,
\item[(iii)] The speed of sound satisfies the bounds $\nu \leq v_s^2\leq \frac{1}{d}$,
\item[(iv)] The quantities $\nu\frac{\eta}{\kappa}$ and $\nu\frac{\zeta}{\kappa}$ are bounded and $\nu\frac{\eta}{\kappa}$ converges to a positive value when $T\to \infty$.
\end{enumerate}
These conditions are automatically fulfilled for a relativistic simple gas of hard spheres or disks (see the previous section). Recall also the condition $c_v < d k_B$ which, together with (i) and (ii), implies that
\begin{equation}
l_0:=\lim\limits_{T\to 0}\frac{v_s^2}{\nu} =
\lim\limits_{T\to 0}\frac{e}{h}\frac{c_p}{c_v}
 = \lim\limits_{T\to 0}\frac{e}{h}\left( 1 + \frac{k_B}{c_v} \right)\geq 1 + \frac{1}{d} > 1.
\label{Eq:l0Limit}
\end{equation}
In particular,  it follows that $v_s\to 0$ for $T\to 0$ since $\nu\to 0$ in this limit. Furthermore, this and Eq. (\ref{Eq:l0Limit}) show that condition (iii) is automatically satisfied for small temperatures. In the opposite limit, using L'H\^opital's rule and condition (ii), one finds
\begin{equation}
l_\infty:=\lim\limits_{T\to\infty}\frac{v_s^2}{\nu} =
\lim\limits_{T\to\infty}\frac{e}{h}\frac{c_p}{c_v}
 =
\lim\limits_{T\to\infty}\frac{c_v}{c_p}\frac{c_p}{c_v} = 1.
\label{Eq:linfLimit}
\end{equation}

After these remarks we are ready to prove the claim. We first note that the coefficients $d_i$ and $c_i$ are positive. Next, using Eqs.~(\ref{Eq:y0},\ref{Eq:yprime0}) it is simple to verify that
\begin{equation}
0 < -y'(0) = \frac{D_2}{2c_1(H_0 H_1)^2},
\end{equation}
such that the inequality~(\ref{Eq:Ineq2}) is automatically satisfied. Consequently, $B_2 > 0$, and hence it only remains to verify that $B_1$ and $B_3$ obey the inequalities $B_1 > 0$ and (\ref{Eq:Ineq1},\ref{Eq:Ineq3},\ref{Eq:Ineq4}). In a next step, we introduce the quantities
\begin{equation}
g_1 := g_3 + \nu\frac{\zeta}{\kappa},
\qquad
g_3 := \frac{k_B}{c_v} - \frac{1}{d},
\end{equation}
which are bounded according to assumptions (ii) and (iv) and in terms of which
\begin{equation}
B_1 = H_0\left( c_3 - v_s^2 - \frac{g_1}{\Lambda} \right),\qquad
B_3 = \nu H_0\left( \frac{1}{d} - \frac{g_3}{\Lambda} \right).
\end{equation}
In particular, it follows that $B_3 > 0$ for large enough values of $\Lambda_0$. The positivity of $B_1$ is then a consequence of $D_3 > 0$. Using $c_2\geq H_0 v_s^2$ one finds
\begin{widetext}
 \begin{eqnarray}
    D_1 & \geq &
 H_0^2\left[ P + \frac{1}{\Lambda}\left( \nu g_3 - \frac{\nu}{d} - v_s^2 g_1 \right) + \frac{\nu g_3}{\Lambda^2}\right],
\\
D_3 & = & 
 \frac{H_0^2\nu}{d}\left[ P +
 \frac{(c_3 -2v_s^2)g_1 + 2\nu g_3 - c_3(c_3 - v_s^2)g_3d}{\Lambda}
  + \frac{dg_3(c_3 g_1 - \nu g_3) - g_1^2}{\Lambda^2}
 \right],
\\
D_4 &\geq & \frac{H_0^4}{\Lambda}\left[v_s^2 P + \frac{\nu v_s^2 \left(g_3 - \frac{2}{d}\right) - 2\nu\left(g_3 + \frac{1}{d}\right)(c_3 - v_s^2) - v_s^4 g_1}{\Lambda} + 2\nu\frac{ g_1\left(g_3 + \frac{1}{d}\right) +  v_s^2 g_3}{\Lambda^2} \right],
 \end{eqnarray}
\end{widetext}  
where
\begin{equation}
P:=v_s^2(c_3 - v_s^2) - \frac{\nu}{d} = \left( \frac{1}{d} - v_s^2 \right)(v_s^2 - \nu) + 2v_s^2\left(1 - \frac{1}{d}\right)\nu\frac{\eta}{\kappa}.
\end{equation}
According to the assumptions, $P > 0$ for all $T > 0$. Furthermore, using the limits~(\ref{Eq:l0Limit},\ref{Eq:linfLimit}) one finds
\begin{align}
    \lim\limits_{T\to 0}\frac{P}{\nu} & \geq \frac{1}{d}(l_0 - 1) > 0,\nonumber \\
\lim\limits_{T\to\infty}\frac{P}{\nu} & = 2\left( 1 - \frac{1}{d} \right)\lim\limits_{T\to \infty}\nu\frac{\eta}{\kappa} > 0.
\end{align}
Therefore, there exists a constant $\delta > 0$ such that $P\geq \delta\nu$ for all $T > 0$. Since the quantities $\nu$, $v_s^2$, $c_3$, $g_1$, $g_3$ and $\nu/v_s^2$ are bounded, there are constants $C_{ij}\geq 0$ such that
\begin{equation}
D_i \geq f_i\left[ \delta - \frac{C_{i1}}{\Lambda_0} - \frac{C_{i2}}{\Lambda_0^2} \right],\qquad
i=1,3,4,
\end{equation}
with $f_1 = H_0^2\nu$, $f_3 = H_0^2\nu^2/d$ and $f_4 = H_0^4 v_s^2\nu/\Lambda$. Hence, by choosing $\Lambda_0$ sufficiently large one can guarantee that $D_i > 0$ for all $i=1,3,4$ and the claim is proven.

Inverting the relation~(\ref{Eq:LambdaGamma1}) and substituting $\Lambda=\Lambda_0/\nu$ leads to Eq.~(36).

\bibliography{refs_kinetic}

\end{document}